\documentclass[preprint,showpacs,preprintnumbers,amsmath,amssymb]{revtex4}
\usepackage{graphicx}
\usepackage{dcolumn}
\usepackage{bm}

\begin{document}

\title{Observation of recoil-induced resonances and electromagnetically induced absorption of cold atoms in diffuse light}

\author{Wen-Zhuo Zhang$^{1,2}$}
\author{Hua-Dong Cheng$^{1}$}
\author{Liang Liu$^{1,3,}$\footnote{Corresponding author:
liang.liu@siom.ac.cn}}
\author{Yu-Zhu Wang$^1$}

\affiliation{$^1$Key Laboratory of Quantum Optics, Shanghai
Institute of Optics and Fine Mechanics, Chinese Academy of
Sciences, Shanghai 201800, China\\
$^2$Graduate University of the Chinese Academy of Sciences,
Beijing 100039, China \\
$^3$State Key Laboratory of Precision Spectroscopy, East China
Normal University, Shanghai 200062, China}

\date{\today}

\begin{abstract}
In this paper we report an experiment on the observation of the
recoil-induced resonances (RIR) and electromagnetically induced
absorption (EIA) of cold $^{87}$Rb atoms in diffuse light. The pump
light of the RIR and the EIA comes from the diffuse light in an
integrating sphere, which also serves the cooling light. We measured
the RIR and the EIA signal varying with the detuning of the diffuse
laser light, and also measured the number and the temperature of the
cold atoms at the different detunings. The mechanism of RIR and EIA
in the configuration with diffuse-light pumping and laser probing
are discussed, and the difference between the nonlinear spectra of
cold atoms in a diffuse-light cooling system and in a
magneto-optical trap (MOT) are studied.
\end{abstract}

\maketitle

Laser spectroscopy of cold atoms is a widely-studied subject in
atom-light interaction due to the negligible Doppler broadening of
cold atoms. With such a feature, laser spectroscopy of cold atoms
is widely applied in studying and manipulating of coherent states
of atoms and light, quantum information processing, as well as
cold atom clocks.

Nonlinear spectra of cold atoms is an important subject in laser
spectroscopy of cold atoms which are usually studied in pump-probe
configuration. In this configuration, a strong light plays the
role of both cooling and pump light. Atoms are cooled to ultra-low
temperature by the strong light and also pumped by it. A weak
probe laser passes through the cold atom cloud to obtain the
nonlinear transmission spectra. Nonlinear spectroscopy of cold
atoms in optical molasses as well as magneto-optical trap (MOT)
has been widely studied \cite{prl91,epl91,pra93,prl93,oc97,pra05},
where the counter-propagating laser beams plays the role of both
cooling and pump light.

Recently, laser cooling of atoms in diffuse light has shown a
great potential in many applications due to its unique features.
In diffuse cooling, laser beams do not need any careful alignment
or collimation and is therefore very robust. Diffuse cooling is an
all-optical cooling technique, and thus has important applications
in cold atom clock \cite{horace01}. Diffuse cooling has a
relatively larger velocity cooling range, and can capture more
atoms than optical molasses or MOT. The first successful
realization of the diffuse cooling directly from atomic vapor was
in cesium atoms \cite{3dcooling}, and followed in rubidium atoms
\cite{Cheng_PRA_09}.

In diffuse light, an atom with velocity $\vec{v}$ can resonate with
the photons from the diffuse laser light, whose propagating
directions distribute on an pyramidal surface which have the same
angle $\theta$ with respect to $\vec{v}$, and $\theta$ and $\vec{v}$
satisfies the resonate condition:
\begin{equation}
\Delta-\vec{k}\cdot\vec{v}\cos{\theta}=0.
\end{equation}

\begin{figure}
\centerline{\includegraphics[width=3in]{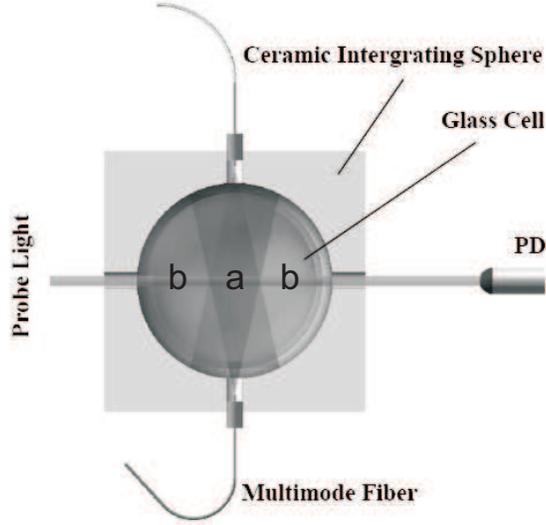}}
\caption{Experimental setup of the diffuse cooling of $^{87}$Rb
atoms. Here, light intensity in the region a is higher than
region b.\label{setup}}
\end{figure}

Although nonlinear spectra of cold atoms are widely studied in the
optical molasses as well as in the MOT, they have been rarely
studied in diffuse laser cooling system. In this paper, we report
an experiment on the observation of the nonlinear spectra of cold
$^{87}$Rb atoms in the diffuse laser light, including the
recoil-induced resonances (RIR) and the electromagnetically
induced absorption (EIA).

Our experimental setup is shown in Fig. \ref{setup}. The diffuse
laser light is created by a ceramic integrating sphere via
Lambertian reflection of two laser beams from two multi-mode
optical fibers. The reflectance of the inner surface of the
ceramic integrating sphere is about $98\%$ for the 780nm light. A
spherical glass cell connected to an ion pump is set inside the
integrating sphere. Inner diameter of the integrating sphere is
$48$ mm and the diameter of the spherical glass cell is $45$ mm.
Vacuum in the glass cell is about 10$^{-9}$ Torr. $^{87}$Rb atomic
vapor is filled in the spherical glass cell and is cooled by the
diffuse laser light \cite{Cheng_PRA_09}.

The cooling laser is supplied by a Toptica TA100 laser system with
total output power of $\sim$ 100 mW and line-width smaller than 1
MHz that is detuned red of the transition of
$5^2$S$_{1/2},F=2\rightarrow 5^2$P$_{3/2},F'=3$ of $^{87}$Rb atom.
A very weak linearly-polarized probe beam of $\sim 1 \mu$W is
split from the cooling laser. Such an arrangement keeps the phase
between cooling beam and probe beam highly correlated, regardless
of the change of environment, such as vibration. A weak repumping
laser with total power of 3.8 mW, which is supplied by a Toptica
DL100 laser system, is mixed into the cooling beam with a
polarizing beam splitter. Frequency of the repumping laser is
locked to the transition between $5^2$S$_{1/2},F=1$ and
$5^2$P$_{3/2},F'=2$. The cooling beam and the repumping laser beam
are injected into the integrating sphere vertically through two
multi-mode fibers. Inside the integrating sphere, the cooling and
repumping beams become the diffuse laser light by Lambertian
reflection. The weak linearly-polarized probe beam
($\sim20\mu$W/cm$^2$) propagates through the center of the
integrating sphere horizontally to obtain the transmission signal.

\begin{figure}
\centerline{\includegraphics[width=4in]{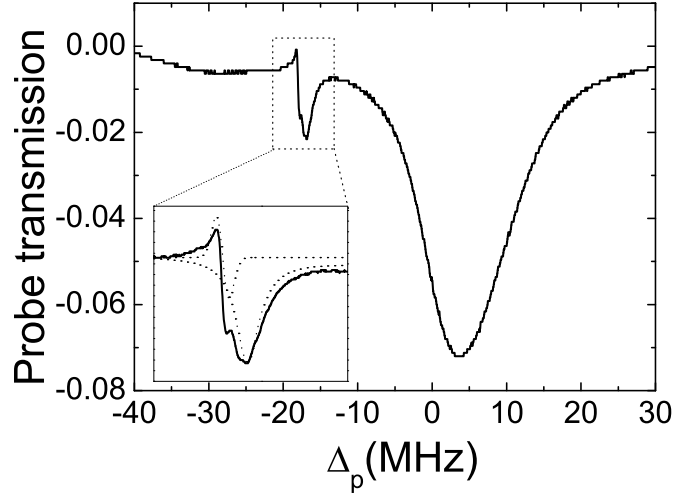}}
\caption{Experimental probe transmission signal of $^{87}$Rb cold
atoms. Here $\Delta_p$ is the detuning of the probe beam. Total
power of the two injected cooling laser beams into the integrating
sphere is $\sim 36$ mW, and their detuning is $\Delta_c=-3.0\Gamma$.
The intensity of the probe laser beam is $\sim $ 20 $\mu$ W/cm$^2$.
The strong absorption signal comes from the linear absorption of
$F=2\rightarrow F'=3$ transition. The weak amplification and
absorption signal near $\Delta_p=-3.0\Gamma\approx-18$MHz is the
nonlinear spectrum signal, which can be decomposed into a derivative
signal and a pure absorption signal (dotted lines).
\label{fix_cooling} }
\end{figure}

We first set the detuning of the cooling laser
$\Delta_c=\omega_c-\omega_0=-3.0\Gamma$ with respect to the
cooling transition. Here $\omega_c$ is the frequency of cooling
laser, $\omega_0$ is the transition frequency between
$5^2$S$_{1/2},F=2$ and $5^2$P$_{3/2},F'=3$, and $\Gamma=6.056$ MHz
is the decay rate of the level $5^2$P$_{3/2},F=3$. Detuning of the
probe laser $\Delta_p=\omega_p-\omega_0$ is swept from
$-7.0\Gamma$ to $6.0\Gamma$ with respect to the cooling transition
by an AOM, where $\omega_p$ is the frequency of the probe laser.
The probe transmission signal is shown in figure
\ref{fix_cooling}, which includes the absorption of
$5^2$S$_{1/2},F=2\rightarrow 5^2$P$_{3/2},F'=3$ transition near
$\Delta_p=0$ and nonlinear spectra near
$\Delta_p=\Delta_c=-3\Gamma$.

\begin{figure}
\centerline{\includegraphics[width=4in]{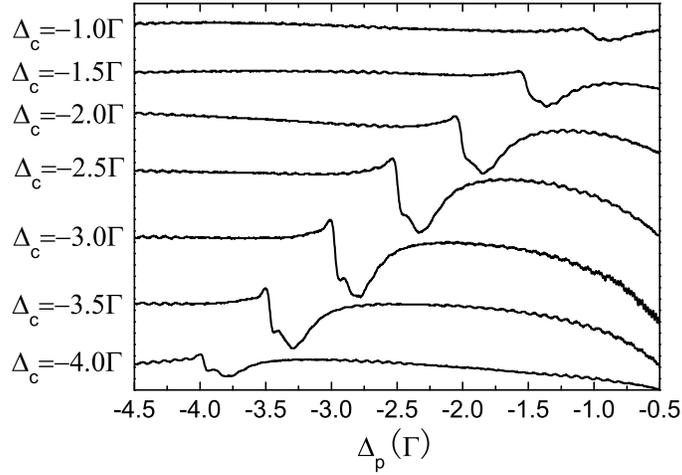}}
\caption{Experimental signal of the nonlinear spectra of cold
atoms varying with the detuning of diffuse laser light
($\Delta_c$). Detuning of the probe laser beam ($\Delta_p$) is
swept from $-4.5\Gamma$ to $-0.5\Gamma$. Total power of the two
injected cooling laser beams into the integrating sphere is $\sim
36$ mW, and the intensity of the probe laser beam is $\sim$20
$\mu$W/cm$^2$. \label{vary_cooling}}
\end{figure}

\begin{figure}
\centerline{\includegraphics[width=4in]{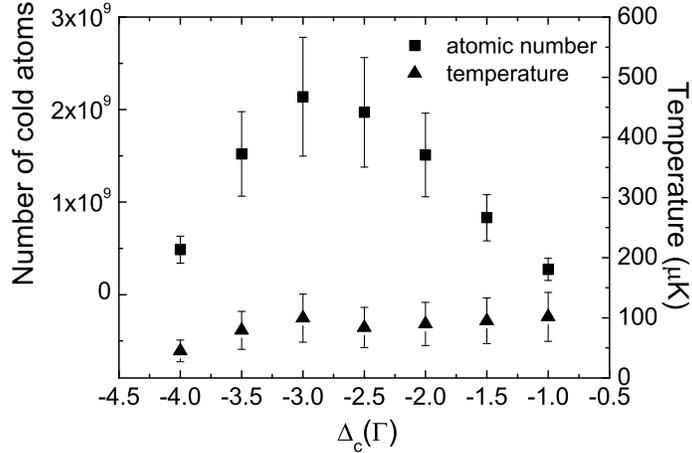}}
\caption{Number and temperature of cold atoms for the various
detunings of cooling laser $\Delta_c$ corresponding to Fig.
\ref{vary_cooling}. Total power of the two injected cooling laser
beams into the integrating sphere is $\sim 36$ mW, and the
intensity of the probe laser beam is $\sim20$
$\mu$W/cm$^2$.\label{number_T}}
\end{figure}

With the power of the cooling laser lights ($36$ mW) and the
intensity of the probe laser beam ($\sim$20 $\mu$W/cm$^2$) stable,
we change the $\Delta_c$ to other values to measure the nonlinear
spectra as well as the number and the temperature of cold atoms
varying with $\Delta_c$. Figure \ref{vary_cooling} shows the
signal of the nonlinear spectra vs $\Delta_p$ with different
$\Delta_c$. The detuning of the probe laser $\Delta_p$ is swept
from $-4.5\Gamma$ to $-0.5\Gamma$. Figure \ref{number_T} gives the
number and the temperature of cold atoms vs $\Delta_c$, which are
measured with the same method used in our previous work
\cite{Cheng_PRA_09}. The temperature is measured from the TOF
(time of flight) signal when the diffuse light is switched off. It
requires that the probe laser beam is horizontally configured.
>From Fig. \ref{vary_cooling} and Fig. \ref{number_T}, we see that
the amplitude of the nonlinear spectrum signal is proportional to
the number of cold atom when the intensity of diffuse light and
the probe light are fixed. The temperature values are below
Doppler limit of the $^{87}$Rb atom, which imply some sub-Doppler
cooling process may happen in our experimental configuration.

Figure \ref{vary_cooling} shows that the signal of nonlinear
spectra appears always when the frequency of probe laser is near
the frequency of cooling laser (the position that
$\Delta_p\approx\Delta_c<0$), and when $\omega_p-\omega_c<0$, the
probe beam obtains a small amplification, whereas when
$\omega_p-\omega_c>0$, the probe beam is absorbed. The
transmission signal of the probe beam is a sum of a derivative
signal and a pure absorption signal. We can see that the
transmission signal has a trend to separate the derivative signal
from the pure absorption signal when $\Delta_c$ becomes a larger
value.

\begin{figure}
\centerline{\includegraphics[width=3in]{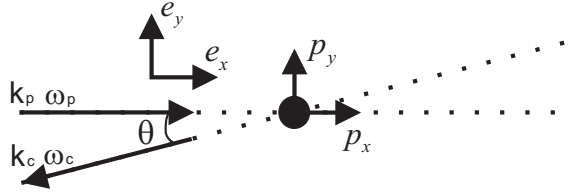}}
\caption{Scheme of recoil-induced resonance. Probe laser
($k_p,\omega_p$) travels along the direction $e_x$, The other beam
($k_c,\omega_c$) is one beam of the isotropic laser light which can
cause recoil-induced resonances of the atom with the probe
laser.\label{scheme_RIR}}
\end{figure}

The derivative signal comes from the RIR. It is the derivative of
the line-shape related to velocity distribution of cold atoms.
This spectrum was first theoretically predicted by Guo \emph{et
al} \cite{recoil92} and was experimentally observed in optical
molasses \cite{recoil94,recoil05}. Figure \ref{scheme_RIR} shows
the scheme of RIR which happens only when the pump and probe beam
are counter-propagating and the angle $\theta$ between the pump
and the probe beam is very small. This is because the RIR is a
two-photon process with stimulated absorbtion of a photon from
pump/probe beam and stimulated emission of a photon to probe/pump
beam. Thus the wave-vector of photon from pump beam ($k_c$), probe
beam ($k_p$), the momentum of atom on $x$ direction ($p_x$) and
$y$ direction ($p_y$) must obey the momentum and energy
conservation, which gives the constraint condition
\begin{equation}
\begin{split}
4\pi m(\omega_p-\omega_c)=&2k_p p_x+2k_c(p_x\cos\theta-p_y\sin\theta)\\
&+\hbar(k_p^2+k_c^2+2k_pk_c\cos\theta),
\end{split}
\end{equation}
where $m$ is the mass of the atom. Because $p_x$ and $p_y$ have
the same Maxwell-Boltzman distribution, it can be proved from the
constraint condition that when $\theta$ is large, the absorption
and amplification of the probe beam can attenuate each other's
signal strength (when $\theta=90^{\circ}$ they can totally cancel
with each other). Only when $\theta$ is close to zero does the
amplification effect on probe beam become dominate at
$\omega_p<\omega_c$ and the absorption effect become dominate at
$\omega_p>\omega_c$. Because of the small-angle condition, the
main contribution from diffuse light in the integrating sphere to
the recoil-induced resonances is the light whose angles to the
probe beam are small. That means the light before first-time
reflection contributes little to the RIR signal. Another method
can also be used to measure the temperature of cold atoms directly
from the peak-to-peak separation of the RIR signal
\cite{velocimetry}, but such method can only obtain the
temperature of the cold atoms which distribute within the probe
beam.  In our experiment, as shown in Fig. \ref{fix_cooling}, the
two peaks of the RIR signal are separated by $\sim$ 500 kHz,
corresponding to the temperature of 80 $\mu$K, which is within the
error range of the measured result of TOF signal in Fig.
\ref{number_T}.

The pure absorption part in the nonlinear spectra is the signal of
EIA \cite{eia99,eia98}, which requires that both of the ground and
excited states have Zeeman sub-levels and the quantum number
$F_e>F_g$. Diffuse pump light is quite suitable for the transfer of
coherence (TOC) in the EIA to happen \cite{TOC}. It is because the
integrating sphere can randomize the polarization of diffuse laser
light \cite{depolarizer}, then the cold atoms can be pumped with all
possible transitions by $\sigma^+$, $\sigma^-$ and $\pi$ lights.
Here the absorption peak of the EIA signal is not on the exact
position of  $\omega_p=\omega_c$ due to the light shift. The FWHM of
the EIA signal generally equals to the decay rate due to time of
flight in the probe beam, therefore with the same beam size of the
probe light the FWHM should be smaller in cold atoms than in
room-temperature atoms due to the much longer interaction time of
cold atoms with the probe beam. However, in the experiment with
room-temperature atoms there is buffer gas which is mixed into the
atomic vapor to make atoms stay  longer time in the probe area, so
very narrow FWHM of EIA signal with 100 kHz order in the
room-temperature can be observed \cite{eia99,eia98}. The EIA can
easily happens no matter what the angle between the pump and probe
beam is, so through the light path of probe beam all the beams in
the diffuse light that intersect with the probe beam can contribute
to the EIA, including the two expanded beams before diffuse
reflection.

Diffuse cooling is a laser cooling method besides the MOT, therefore
it is interesting to compare the nonlinear spectra between them. The
main difference between the diffuse laser and the
$\sigma^+$$\sigma^-$ configured three-dimensional optical molasses
of a MOT is the polarization of the light field. Cold atoms can have
all $\Delta m_{F}=0,\pm 1$ transitions corresponding to $\pi$,
$\sigma^+$, and $\sigma^-$ polarization of the pump light. The light
field of a MOT is a three-dimensional optical molasses, where the
polarization of each beam at every position is fixed. We can know
the exact polarization at every position of the light field if we
know the exact polarization of each beam at each position. However,
diffuse light is created as well as depolarized by the Lambertian
inner-surface of the integrating sphere \cite{depolarizer}, so it is
a field with random polarization directions, and the polarization
distributes randomly over the space.

Because the polarization at every position of the light field in
the center of the MOT is known, the transition of the trapped cold
atoms can be predicted if the direction of the magnetic field
$\vec{B}(x)$ is also known at every position. A well trapped cold
atom may experience periodic light polarization and magnetic
field, which may cause different light-shift and steady-state
population among every ground state Zeeman sub-levels. This case
has been studied by Brzozowski \emph{et al} \cite{pra05}. They
theoretically predicted that the weight of the $\pi$ transition of
the cold atom stands out, which makes the population weight of the
$m_F=0$ ground-state Zeeman sub-level more than others and
suitable for the stimulated Raman process
\cite{prl91,epl91,pra93,prl93,oc97,pra05}.

Contrarily, in the diffuse laser light the polarization is totally
randomized. Since the diffuse light has only cooling but no trapping
effect on atoms, the cold atoms are not well trapped. Therefore a
cold atom in the diffuse laser light can be pumped randomly at every
position with all $\Delta m_{F}=0,\pm 1$ transitions no matter what
the direction of the quantization axis (the direction of an strong
magnetic field $\vec{B}(x)$) is. This feature makes it difficult to
have significant population difference among all ground state Zeeman
sub-levels, which limits stimulated Raman process but quite suitable
for the EIA. The random polarized pump light can easily have
different polarizations with the probe light naturally, which is a
necessary condition for EIA-TOC \cite{TOC}. Then it is more likely
to observe the EIA-TOC of cold atoms in diffuse laser light.

In conclusions, we have observed the signal of recoil-induced
resonances (RIR) and electromagnetic-induced absorption (EIA) of
cold $^{87}$Rb atoms in an integrating sphere, where the atoms are
cooled and pumped by the diffuse laser light. We analyzed the
mechanism of nonlinear spectra of cold atoms in diffuse laser light,
and show its differences to the case in MOT. The simple experimental
setup and the unique feature of nonlinear spectra of cold atoms in
diffuse laser light make the subject is worth to study for the
future.

This work is supported by the National Nature Science Foundation of
China under Grant No. 10604057 and National High-Tech Programme
under Grant No. 2006AA12Z311.

\end{document}